%% file: main.tex
\documentclass[preprint,a4paper,12pt]{elsarticle}
\pdfoutput=1
\usepackage{amsmath,amssymb,graphicx,todonotes,xspace,hyperref,slashed,subcaption,float,tabulary}
\usepackage{booktabs}
\usepackage{listings}
\usepackage{multirow}
\usepackage[absolute]{textpos}
\usepackage[section]{placeins}
\hypersetup{
  pdftitle={HEJ 2.2},
  colorlinks,
  urlcolor={blue},
  linkcolor={blue},
}

\graphicspath{{figures/}{build/figures/}}

\lstdefinelanguage{sherpa}{%
  morekeywords={run,processes,selector,analysis},
  sensitive=true,
  morecomment=[l]{\#},
}

\lstdefinelanguage{yaml}{%
  morekeywords={true,false},
  sensitive=false,
  morecomment=[l]{\#},
}

\newcommand{\HEJ}{{\tt HEJ}\xspace}
\newcommand{\HEJFOG}{{\tt HEJFOG}\xspace}

\newcommand{\HIGHEJ}{\emph{High Energy Jets}\xspace}

\newcounter{bla}

\begin{document}

\begin{frontmatter}
\title{\HEJ 2.2: W boson pairs and Higgs boson plus jet production at high energies}

\author[a]{Jeppe~R.~Andersen}
\author[b]{Bertrand~Duclou\'e}
\author[b]{Conor~Elrick}
\author[a]{Hitham~Hassan}
\author[c]{Andreas~Maier}
\author[b]{Graeme~Nail}
\author[b]{J\'er\'emy~Paltrinieri}
\author[d]{Andreas~Papaefstathiou}
\author[b]{Jennifer~M.~Smillie}

\address[a]{Institute for Particle Physics Phenomenology, University of Durham,
Durham, DH1 3LE, UK}
\address[b]{Higgs Centre for Theoretical Physics, University of Edinburgh,
  Peter Guthrie Tait Road, Edinburgh, EH9 3FD, UK}
\address[c]{Deutsches Elektronen-Synchrotron DESY, Platanenallee 6,
  15738 Zeuthen, Germany}
\address[d]{Department of Physics, Kennesaw State University, Kennesaw, GA 30144, USA}

\begin{abstract}
We present version 2.2 of the \HIGHEJ (\HEJ) Monte Carlo event
generator for hadronic scattering processes at high energies. The new
version adds support for two further processes of central
phenomenological interest, namely the production of a W boson pair
with equal charge together with two or more jets and the production of
a Higgs boson with at least one jet.
Furthermore, a new prediction for charged lepton pair production with
high jet multiplicities is provided in the high-energy limit. The
accuracy of \HEJ 2.2 can be increased further through an enhanced
interface to standard predictions based on conventional perturbation
theory. We describe all improvements and
provide extensive usage examples. \HEJ 2.2 can be obtained from
\url{https://hej.hepforge.org}.
\end{abstract}

\begin{keyword}
  Collider Physics; Monte Carlo Event Generation; Resummation
\end{keyword}

\begin{textblock*}{10cm}(100mm,22mm)
IPPP/23/18, DCPT/23/36, DESY-23-038
\end{textblock*}
\end{frontmatter}

{\bf NEW VERSION PROGRAM SUMMARY}

\begin{small}
\noindent
{\em Program Title:} \HEJ.                                          \\
{\em Licensing provisions:} GPLv2 or later.\\
{\em Programming language:} C++.                               \\
{\em Journal reference of previous version:} Comput.Phys.Commun. 278 (2022) 108404.\\
{\em Does the new version supersede the previous version?:} Yes.  \\
{\em Reasons for the new version:} Support for further scattering processes and improved combination with fixed-order predictions.\\
{\em Summary of revisions:} The new release adds the ability to predict the QCD
component in the high-energy production of two leptonically decaying W bosons with
equal charge, together with two or more jets. High-energy resummation
is now also implemented for the production of a Higgs boson together
with a single jet, whereas before only processes involving at least
two jets had been considered. Pure resummed predictions for lepton
pair production via a virtual photon or Z boson together with jets are
now provided for high jet multiplicities, where fixed-order matching is
no longer feasible. The interface to fixed-order generators has been
extended significantly, including options for differential reweighting
to next-to-leading order, filtering of jets with low transverse
momentum, and the capability to stream a wider range of input event
files.  \\
{\em Nature of problem:}
Hadronic scattering processes at high energies, i.e.~with large invariant
masses between jets, are of great phenomenological interest. This is
in particular the case for measurements of weak-boson scattering and
weak-boson fusion production of a Higgs boson. In the high-energy
region, standard perturbation theory exhibits poor convergence for the QCD
contributions, which limits the predictive power of conventional Monte
Carlo generators.\\
{\em Solution method:}
The poor convergence of the perturbative series can be traced to the
appearance of large high-energy logarithms. \HEJ is a fully flexible
event generator combining fixed-order accuracy with the all-order
resummation of such logarithms, based on the \HIGHEJ framework.
The new version \HEJ 2.2 provides accurate predictions
for a range of processes of central phenomenological interest,
including the QCD component of same-sign W boson pair production with multiple jets and Higgs
boson production in association with one or more jets.
\end{small}

\newpage
\tableofcontents

\input{Introduction}
\input{Features}
\input{Application}
\input{Conclusions}

\newpage
\appendix
\input{Installation}

\bibliographystyle{elsarticle-num}
\bibliography{papers}

\end{document}

%% file: Introduction.tex
\section{Introduction}
\label{sec:intro}

Hadronic scattering processes in the high-energy region are of great
phenomenological interest. Prime examples include coupling
measurements in weak boson fusion and weak boson scattering.  To
suppress the background in these measurements, one typically requires
jets with large invariant masses and a large difference in
rapidity. These requirements strongly enhance the contribution from
the high-energy region, which is characterised by invariant masses
that are much larger than all transverse scales, or equivalently large
rapidity separations with no strong hierarchy between the transverse
momenta. In this kinematic regime, high-energy logarithms of the large ratio $\hat{s}
/p_\perp^2$ arise in perturbation theory, where $\hat{s}$ is
the square of the partonic centre-of-mass energy and $p_\perp$ the characteristic
transverse momentum scale. In the QCD component, these large
logarithms jeopardise the convergence of the perturbative series.

\HIGHEJ (\HEJ) is both a framework and a flexible Monte Carlo
generator for the all-order resummation of high-energy
logarithms~\cite{Andersen:2009nu,Andersen:2009he,Andersen:2011hs}. In
\HEJ 2~\cite{Andersen:2019yzo}, this high-energy resummation can
additionally be matched to leading-order predictions obtained using
conventional fixed-order generators~\cite{Andersen:2017kfc}. \HEJ has
been validated against data in experimental studies of pure multijet
production~\cite{ATLAS:2011yyh,CMS:2012rfo,CMS:2012xfg,ATLAS:2014lzu},
lepton pair production via a virtual W boson, photon, or Z boson in
association with two or more
jets~\cite{Andersen:2012gk,D0:2013gro,ATLAS:2014fjg,Andersen:2016vkp},
and Higgs boson production with jets~\cite{Andersen:2022zte}.

In the following, we present \HEJ 2.2. This new version implements
high-energy resummation for the production of two leptonically
decaying W bosons with the same charge in association with two or more
jets. Moreover, the existing implementation for the production of a
Higgs boson with jets has been extended to also cover Higgs boson
production with a single jet. Resummation for the production of a
charged lepton pair via a virtual photon or Z boson together with two
or more jets is now also supported for higher multiplicities where no
fixed-order prediction is available. Furthermore, new options have
been added, for instance to facilitate differential
next-to-leading-order matching and to separate events with soft
jets. In section~\ref{sec:features}, we briefly summarise \HIGHEJ and
describe the various improvements in version 2.2. We give examples for
the use of the new features in section~\ref{sec:usage} and conclude in
section~\ref{sec:conclusions}. \ref{sec:installation}
contains instructions for download and installation of \HEJ 2.2.


%% file: Features.tex
\section{Features of \HEJ 2.2}
\label{sec:features}

\subsection{\HEJ in a nutshell}
\label{sec:HEJ_summary}

Before describing the changes in \HEJ 2.2, let us briefly review the
general formalism and program structure. As input, \HEJ requires
leading-order (LO) events, generated with
e.g.~\textsc{Sherpa}~\cite{Bothmann:2019yzt} or
\textsc{MadGraph5\_aMC@NLO}~\cite{Alwall:2014hca}. For higher jet
multiplicities exact fixed-order generation becomes increasingly time
consuming. To address this problem, \HEJ includes the fast \HEJ
fixed-order generator \HEJFOG based on the high-energy approximation
of the leading-order matrix elements.

\begin{figure}[htb]
  \centering
  \includegraphics{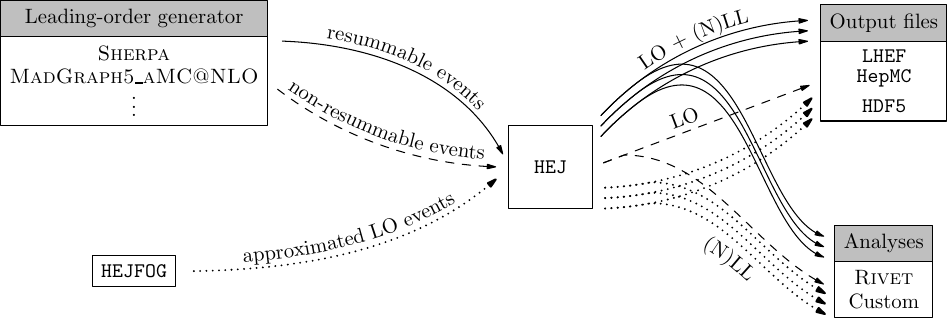}
  \caption{Standard \HEJ control flow.}
  \label{fig:flow}
\end{figure}

Using the kinematics of each (approximate or exact) input event, we
identify whether resummation is possible. For each event that permits
resummation, \HEJ generates a number of matching events in the
resummation phase space, which include real and virtual corrections to
all orders in the high-energy limit. Details are given
in~\cite{Andersen:2018tnm}. Together with the unchanged non-resummable
input events, the generated resummation events are then passed on to
any number of output event files and/or analyses. This standard control flow is
depicted in figure~\ref{fig:flow}. It can be modified through \HEJ
options, such that e.g.~non-resummable events are discarded.

The first type of event kinematics for which resummation is
implemented are leading-logarithmic (LL) configurations, which for
pure multijet production have to fulfil the following constraints:
\begin{enumerate}
\item The flavour of the most backward outgoing parton has to match
the flavour of the backward incoming parton.
\item The flavour of the most forward outgoing parton has to match
the flavour of the forward incoming parton.
\item All other outgoing partons have to be gluons.
\end{enumerate}
These criteria remain the same in processes involving virtual photons
and/or Z bosons. For virtual W bosons, the incoming and outgoing
flavours in criteria 1 and 2 only have to match up to the change
induced by W boson couplings. In the case of a final-state Higgs
boson, configurations where the backward (forward) incoming parton is
a quark or antiquark and the most backward (forward) outgoing particle
is the Higgs boson are formally subleading. Nevertheless, we also
implement resummation for such configurations. Depending on the
process, resummation is also implemented for two further types of
next-to-leading-logarithmic (NLL) configurations. These configurations
differ from LL ones as follows.
\begin{itemize}
\item \emph{Unordered gluon:} Either the most forward or most backward
outgoing parton is a gluon, and the next outgoing parton in rapidity
order is a quark or antiquark whose flavour matches the one of the
respective incoming parton.
\item \emph{Quark-antiquark:} A pair of final-state gluons that are
adjacent in rapidity is replaced by a quark-antiquark pair.
\end{itemize}
The current status of the implemented resummation is summarised in
table~\ref{tab:summary}.

\begin{table}[tb]
  \centering
  \begin{tabular}{lcccc}
    \toprule
    Process & pure LL & LO + LL & \multicolumn{2}{c}{NLL} \\
    \cmidrule(lr){4-5}
    &&&unordered gluon & quark-antiquark\\
    \midrule
    $\geq$2 jets & \HEJ 2.0 & \HEJ 2.0 & \HEJ 2.0 & \HEJ 2.1\\
    H + 1 jet & --- & \HEJ 2.2 & N/A & N/A \\
    H + $\geq$2 jets & \HEJ 2.0 & \HEJ 2.0 & \HEJ 2.0 & ---\\
    W + $\geq$2 jets & \HEJ 2.1 & \HEJ 2.1 & \HEJ 2.1 & \HEJ 2.1\\
    Z/$\gamma$ + $\geq$2 jets & \HEJ 2.2 & \HEJ 2.1 & \HEJ 2.1 & ---\\
    W$^\pm$W$^\pm$ + $\geq$2 jets & --- & \HEJ 2.2 & --- & ---\\
    \bottomrule
  \end{tabular}
  \caption{%
    Implemented processes and higher-order logarithmic corrections in
    \HEJ. The ``pure LL'' column lists processes implemented in the
    \HEJFOG. The NLL columns include both pure NLL and NLL matched to LO.
  }
  \label{tab:summary}
\end{table}

The resummation events generated for the LL and supported NLL
configurations are given a final matrix element weight of
\begin{equation}
  \label{eq:weight}
  \lvert\mathcal{M}_{\HEJ}\rvert^2 \frac{\lvert\mathcal{M}_{\text{LO}}\rvert^2}{\lvert\mathcal{M}_{\HEJ, \text{LO}}\rvert^2},
\end{equation}
where
$\mathcal{M}_{\HEJ}$ is the all-order scattering matrix element in the
high-energy approximation, $\mathcal{M}_{\HEJ, \text{LO}}$ its
leading-order truncation, calculated for the kinematics of the input
event and $\lvert\mathcal{M}_{\text{LO}}\rvert^2$ is taken from the LO input.

To illustrate the structure of the \HEJ matrix element, we first focus
on LL configurations in pure multijet production. We denote these
configurations as $f_a f_b \to f_a \cdots f_b$, where $f_a$ is the
flavour of the incoming parton in the backward direction with momentum
$p_a$. Correspondingly, we use $f_b$ and $p_b$ for the flavour and
momentum of the forward incoming parton. The final state contains $n$
partons with momenta $p_1,\dots,p_n$, which we order by rapidity,
viz.~$y_1 < \dots < y_n$. The most backward outgoing parton has
flavour $f_{a}$, the most forward one flavour $f_{b}$, and all other
outgoing partons are gluons. This implies that the outgoing parton
with flavour $f_a$ has momentum $p_1$ and the outgoing parton with
flavour $f_b$ has momentum $p_n$. However, note that this
identification does not necessarily hold at NLL. For example, if there
is an unordered gluon in the backward direction then the outgoing
parton with flavour $f_a$ has momentum $p_2$. Using the introduced
notation, we can write the general form of the squared \HEJ matrix
element as
\begin{equation}
  \label{eq:ME_fact}
  \begin{split}
    \overline{\left\lvert\mathcal{M}_{\HEJ}^{f_a f_b \to f_a \cdots f_b}\right\rvert}^2 ={}&\mathcal{B}_{f_a,f_b}(p_a, p_b, p_1, p_n)\\
    &\cdot \prod_{i=1}^{n-2} \mathcal{V}(p_a,p_b,p_1,p_n, q_i, q_{i+1})\\
    &\cdot \prod_{i=1}^{n-1} \mathcal{W}(q_{i\perp}, y_i, y_{i+1}),
  \end{split}
\end{equation}
where $q_i=p_a - \sum_{j=1}^i p_j$ is the $t$-channel momentum after
the emission of parton $i$. $\mathcal{B}_{f_a,f_b}$ is derived from the modulus
square of the Born-level matrix element for the process $f_a f_b \to
f_a f_b$, $\mathcal{V}$ accounts for the real emission of the $n-2$
gluons in between $f_a$ and $f_b$, and $\mathcal{W}$ incorporates the
virtual and unresolved real corrections.

The Born-level function $\mathcal{B}_{f_a,f_b}$ is given by
\begin{equation}
  \label{eq:B}
  {\cal B}_{f_a,f_b}(p_a, p_b, p_1, p_n) = \frac{(4\pi\alpha_s)^n}{4 (N_C^2 - 1)} \frac{K_{f_a}}{q_1^2} \frac{K_{f_b}}{q_{n-1}^2} \|S_{f_a f_b \to f_{a}\cdots f_{b}}\|^2 ,
\end{equation}
where $\alpha_s$ is the strong coupling constant and $N_C = 3$ the number of colours. $K_{f_a}$ and $K_{f_b}$ are
generalised colour factors depending on the respective parton flavour
and, in the case of gluons, also the parton momentum. For quarks and
antiquarks one finds $K_f = C_F = \frac{N_C^2-1}{2N_C}$; the factor $K_g$ for gluons is
derived in~\cite{Andersen:2009he}. $S_{f_a f_b \to f_{a}\cdots f_{b}}$
denotes the contraction of two currents:
\begin{equation}
  \label{eq:S_QCD}
  \| S_{f_a f_b \to f_a \cdots f_b}\|^2 \equiv \| j^a \cdot j^b\|^2 = \sum_{\substack{\lambda_a=+,-\\ \lambda_b=+,-}}\lvert j^{\mu,\lambda_a}(p_1, p_a) j_\mu^{\lambda_b}(p_n, p_b)\rvert^2,
\end{equation}
where $j_\mu^\lambda$ is the current
\begin{equation}
  \label{eq:j}
  j_\mu^\lambda(p,q) = \bar{u}^\lambda(p) \gamma_\mu u^\lambda(q)
\end{equation}
for helicity $\lambda$. \HEJ employs the symbolic manipulation
language \texttt{FORM}~\cite{Vermaseren:2000nd} to generate compact
symbolic expressions for current contractions.

The real corrections are given by contractions of Lipatov vertices~\cite{Andersen:2017kfc}:
\begin{align}
  \label{eq:V}
  \mathcal{V}(p_a,p_b,p_1,p_n, q_i, q_{i+1}) ={}& -\frac{C_A}{q_i^2 q_{i+1}^2} V_\mu(p_a,p_b,p_1,p_n, q_i, q_{i+1}) V^\mu(p_a,p_b,p_1,p_n, q_i, q_{i+1}),\\
  \label{eq:V_Lipatov}
  V^\mu(p_a,p_b,p_1,p_n, q_i, q_{i+1})={}& -(q_i+q_{i+1})^\mu \nonumber\\
  &+ \frac{p_a^\mu}{2} \left( \frac{q_i^2}{p_{i+1}\cdot p_a} +
  \frac{p_{i+1}\cdot p_b}{p_a\cdot p_b} + \frac{p_{i+1}\cdot p_n}{p_a\cdot p_n}\right) +
p_a \leftrightarrow p_1 \nonumber\\
  &- \frac{p_b^\mu}{2} \left( \frac{q_{i+1}^2}{p_{i+1} \cdot p_b} + \frac{p_{i+1}\cdot
      p_a}{p_b\cdot p_a} + \frac{p_{i+1}\cdot p_1}{p_b\cdot p_1} \right) - p_b
  \leftrightarrow p_n,
\end{align}
with $C_A = N_C$.

Finally, the virtual and unresolved real corrections $\mathcal{W}$ can
be expressed in terms of the regularised Regge trajectory $\omega^0$:
\begin{equation}
  \label{eq:W}
  \mathcal{W}(q_{j\perp},y_j,y_{j+1}) = \exp[\omega^0(q_{j\perp}) (y_{j+1}- y_j)].
\end{equation}
For a detailed discussion and an explicit expression for $\omega^0$
see~\cite{Andersen:2018tnm}.

The generalisations to NLL configurations and additional non-partonic
final state particles are derived
in~\cite{Andersen:2012gk,Andersen:2016vkp,Andersen:2017kfc,Andersen:2018kjg,Andersen:2020yax,Andersen:2021vnf,Andersen:2022zte}. In
all cases one finds a factorisation into a Born-level function,
resolved real emissions, and virtual and unresolved real
corrections. In the absence of interference, one recovers the same
structure as in equation~\eqref{eq:ME_fact}. In particular,
the functions $\mathcal{V}$ and $\mathcal{W}$ comprising the all-order
corrections are universal, whereas the Born-level function
$\mathcal{B}$ is process dependent.

\subsection{High-energy resummation for W pair production}
\label{sec:HEJ_WW}

Based on the pure-QCD LL configurations $f_a f_b \to f_a \cdots f_b$,
additional W bosons can be produced via emission off the partons $f_a$
and $f_b$. In \HEJ 2.2, we consider LL configurations with two
leptonically decaying W bosons with equal charge. For definiteness, we
discuss configurations
$f_a f_b \to (W^- \to e \bar{\nu}_e) (W^- \to \mu \bar{\nu}_\mu)
f_{a'} \cdots f_{b'}$, where the rapidities of the final-state charged
and neutral leptons do not necessarily respect any rapidity
ordering. Note that the couplings to the W bosons induce flavour
changes $f_a \to f_{a'}$ and $f_b \to f_{b'}$. We use
a diagonal CKM matrix and do not include third-generation quarks/anti-quarks, i.e.\ the number of active flavours is 4.  The
production of two positively charged W bosons and the decay of the two
W bosons into the same lepton flavours is completely analogous.

We identify two contributions to the amplitude. Parton $f_a$ can
either couple to the W boson that decays into an electron and its antineutrino or
to the W boson decaying into a muon and its antineutrino. In the first case, the $t$-channel momenta are given by
\begin{equation}
  \label{eq:t-channel_e}
  q_{i,e} = p_a - p_e - p_{\bar{\nu}_e} - \sum_{j=1}^i p_j,
\end{equation}
where $p_e$ is the momentum of the electron and $p_{\bar{\nu}_e}$ the
momentum of its antineutrino. In the second case the $t$-channel momenta
are
\begin{equation}
  \label{eq:t-channel_mu}
  q_{i,\mu} = p_a - p_\mu - p_{\bar{\nu}_\mu} - \sum_{j=1}^i p_j
\end{equation}
with the muon momentum $p_\mu$ and the corresponding antineutrino momentum
$p_{\bar{\nu}_\mu}$. The resulting modulus square of the matrix element including interference is~\cite{Andersen:2021vnf}
\begin{equation}
\label{eq:ME_Z}
\begin{split}
  \left\lvert\mathcal{M}_{\HEJ}^{f_a f_b\to e\bar{\nu}_e \mu \bar{\nu}_\mu f_{a'} \cdots f_{b'}}\right\rvert^2 &=\ \frac{(4 \pi \alpha_s)^n}{4(N_c^2-1)}\ K_{f_a}K_{f_b} C_A^{n-2}\\
  \times \Bigg( &{\frac{\| j^a_{W,e}\cdot
    j^b_{W,\mu}\|^2}{q^2_{1,e}q^2_{n-1,e}}}
\prod^{n-2}_{i=1} {\frac{-V^2(q_{i,e},
  q_{i+1,e})}{q^2_{i,e} q^2_{i+1,e}}} \prod_{i=1}^{n-1}
  \mathcal{W}(q_{i,e\perp},y_i,y_{i+1})\\
+\ &{\frac{\|j^a_{W,\mu} \cdot j^b_{W,e} \|^2}{q^2_{1,\mu}q^2_{n-1,\mu}}}
     \prod^{n-2}_{i=1}{\frac{-V^2(q_{i,\mu}, q_{i+1,\mu})}{q^2_{i,\mu} q^2_{i+1,\mu}}} \prod_{i=1}^{n-1}
\mathcal{W}(q_{i,\mu\perp},y_i,y_{i+1}) \\
-\ &{\frac{2\Re\{ (j^a_{W,e}\cdot j^b_{W,\mu})(\overline{j^a_{W,\mu} \cdot
      j^b_{W,e}})\}}{\sqrt{q^2_{1,e}q^2_{1,\mu}}\sqrt{q^2_{n-1,e} q^2_{n-1,\mu}}}}\\
&\times \prod^{n-2}_{i=1}{\frac{V(q_{i,e}, q_{i+1,e})\cdot V(q_{i,\mu},
         q_{i+1,\mu})}{\sqrt{q^2_{i,e}q^2_{i,\mu}} \sqrt{q^2_{i+1,e}q^2_{i+1,\mu}}}} \prod_{i=1}^{n-1}
         \mathcal{W}(\sqrt{q_{i,e\perp}q_{i,\mu\perp}},y_i,y_{i+1})\Bigg).
\end{split}
\end{equation}
Here, we have introduced contractions between generalised currents
$j^c_{W,l}$ accounting for the coupling between a parton with flavour
$f_c$ and a W boson decaying into a charged lepton $l$ and the
corresponding antineutrino. The contractions are
\begin{align}
  \label{eq:j_contr_sq_1}
  \| j^a_{W,e} \cdot j^b_{W,\mu}\|^2 ={}& \left\lvert j_V^{\rho,\lambda_a \lambda_e}(p_1, p_a, p_e, p_{\bar{\nu}_e}) j_{V}^{\sigma,\lambda_b\lambda_\mu}(p_{n+2}, p_b, p_\mu, p_{\bar{\nu}_\mu}) g_{\rho\sigma}\right\rvert^2,\\
  \label{eq:j_contr_sq_2}
  \| j^a_{W,\mu} \cdot j^b_{W,e}\|^2 ={}& \left\lvert j_V^{\rho,\lambda_a \lambda_\mu}(p_1, p_a, p_\mu, p_{\bar{\nu}_\mu}) j_{V}^{\sigma,\lambda_b\lambda_e}(p_{n+2}, p_b, p_e, p_{\bar{\nu}_e})g_{\rho\sigma}\right\rvert^2,\\
  \label{eq:j_contr_mixed}
  \begin{split}
  (j^a_{W,e}\cdot j^b_{W,\mu})(\overline{j^a_{W,\mu} \cdot
      j^b_{W,e}})={}&
    j_V^{\rho,\lambda_a \lambda_e}(p_1, p_a, p_e, p_{\bar{\nu}_e}) j_{V}^{\sigma,\lambda_b\lambda_\mu}(p_{n+2}, p_b, p_\mu, p_{\bar{\nu}_\mu}) g_{\rho\sigma} \\
    &\qquad\times \overline{j_{V}^{\alpha\lambda_a \lambda_\mu}(p_1, p_a, p_\mu,
      p_{\bar{\nu}_\mu}) j_V^{\beta,\lambda_b \lambda_e}(p_{n+2}, p_b, p_e,
      p_{\bar{\nu}_e})}g_{\alpha\beta} ,
  \end{split}
\end{align}
where the parton helicities $\lambda_a$ and $\lambda_b$ are determined
by the flavour of the respective parton, namely $\lambda_c=-$ if $f_c$
is a quark and $\lambda_c=+$ if $f_c$ is an antiquark. The electron
helicity $\lambda_e$ and the muon helicity $\lambda_\mu$ correspond to
the charge sign of the parent W boson, i.e.\ $\lambda_e = \lambda_\mu
= -$ in the present case. We have introduced a generalised current
$j_V^{\rho,\lambda_a \lambda_\ell}$ for the coupling of a parton with
helicity $\lambda$ to a leptonically decaying vector
boson\footnote{Here, we assume a W boson. However, the same expression
holds for neutral vector bosons after replacing the antineutrino
momentum $p_{\bar{\nu}_{\ell}}$ by the antilepton momentum
$p_{\bar{\ell}}$.} $V$ with lepton helicity $\lambda_{\ell}$. It is
given by~\cite{Andersen:2020yax}
\begin{equation}
  \label{eq:j_V}
  \begin{split}
  j_V^{\rho,\lambda_a \lambda_\ell}(p_1,p_a,p_{\ell},p_{\bar{\nu}_{\ell}}) =&\ \frac{g_V^2}{2}\
     \frac1{p_V^2-M_V^2+i\ \Gamma_V M_V}\ \bar{u}^{\lambda_{\ell}}(p_\ell) \gamma_\alpha
                                               v^{\lambda_{\ell}}(p_{\bar{\nu}_\ell}) \\
& \cdot \left( \frac{ \bar{u}^{\lambda_a}(p_1) \gamma^\alpha (\slashed{p}_V +
  \slashed{p}_1)\gamma^\rho u^{\lambda_a}(p_a)}{(p_V+p_1)^2} +
\frac{ \bar{u}^{\lambda_a}(p_1)\gamma^\rho (\slashed{p}_a - \slashed{p}_V)\gamma^\alpha u^{\lambda_a}(p_a)}{(p_a-p_V)^2} \right).
\end{split}
\end{equation}
$p_V=p_{\ell}+p_{\bar{\nu}_\ell}$ is the vector boson momentum, $g_V$ its
coupling to the fermion $f_a$, $M_V$ its mass, and $\Gamma_V$ the
width.

\subsection{Higgs boson production with a single jet}
\label{sec:HEJ_Hj}

In the gluon-fusion production of a Higgs boson together with one or
more jets new LL configurations beyond those derived from pure
multijet production (c.f.~section~\ref{sec:HEJ_summary}) arise. In
these configurations, one of the incoming partons is a gluon while the
corresponding most forward or most backward outgoing particle is the
Higgs boson, i.e.~$g f_b \to H\cdots f_b$ or $f_a g \to f_a \cdots
H$. Without loss of generality we consider the former
configuration. The modulus square of the \HEJ matrix element reads~\cite{Andersen:2022zte}
\begin{align}
  \label{eq:ME_fact_outer}
  \begin{split}
    \overline{\left\lvert\mathcal{M}_{\HEJ}^{g f_b \to H \cdots f_b}\right\rvert}^2 ={}&\mathcal{B}_{H, f_b}(p_a, p_b, p_1, p_n)\\
    &\cdot \prod_{i=1}^{n-2} \mathcal{V}(p_a, p_b, p_a, p_n, q_i, q_{i+1})\\
    &\cdot \prod_{i=1}^{n-1} \mathcal{W}(q_{i\perp}, y_i, y_{i+1}),
  \end{split}
\end{align}
with the universal real and virtual correction factors $\mathcal{V}$
and $\mathcal{W}$ defined in equations~\eqref{eq:V} and
\eqref{eq:W}. The only differences to the pure QCD case in
equation~\eqref{eq:ME_fact} are the replacement $p_1 \to p_a$ in the
third argument of $\mathcal{V}$ and the adjustment of the
process-dependent Born function to~\cite{Andersen:2022zte}
\begin{align}
  \label{eq:B_Hq}
  \mathcal{B}_{H,f_b} ={}& \frac {(4\pi\alpha_s)^{n-1}} {4(N_c^2-1)}
                           \frac {1} {q_1^2}
                           \frac{K_{f_b}}{q_{n-1}^2}
                           \left\|S_{g f_b \to H f_b}\right\|^2,\\
  \label{eq:S_Hf}
  \left\|S_{g f_b \to  H f_b}\right\|^2 ={}&  \sum_{
  \substack{
  \lambda_{a} = +,-\\
  \lambda_{b} = +,-
  }}
 \left\lvert\epsilon_\mu^{\lambda_a}(p_a)\ V_H^{\mu\nu}(p_a, p_a-p_1)\ j_\nu^{\lambda_b}(p_n, p_b)\right\rvert^2,
\end{align}
where $\epsilon^{\lambda_a}(p_a)$ is the polarisation vector of the
incoming gluon and $V_H$ the effective vertex coupling the Higgs boson to two
gluons at one-loop, including finite quark-mass dependence.  The structure of
equation~(\ref{eq:ME_fact_outer}) then allows finite quark-mass dependence to be
applied for arbitrarily high numbers of jets.

\subsection{Spillover from small transverse momenta}
\label{sec:low_pt}

As described in section~\ref{sec:HEJ_summary}, a number of all-order
resummation events are generated for each resummable input event.
Since the resummation events include real corrections, the resulting
kinematics differ slightly from the kinematics of the input
events. While jet rapidities are always preserved, this is
generally neither true for transverse momenta nor for the rapidities
of any other particles. This implies that cuts imposed on the
leading-order generation should be significantly looser than the final
analysis cuts. Empirically, the difference in transverse momentum cuts
should be about 10-20\%, with a slight increase towards larger
multiplicities.

Simply adjusting the cuts in the leading-order generation is correct,
but inefficient: events with small transverse momenta dominate the
leading-order prediction, but only give a small contribution to the
final resummed results. It is therefore more efficient to split up the leading-order
generation. One first generates a high-statistics sample in which all
particles fulfil the transverse momentum cuts of the analysis. Then,
one generates a second low-statistics sample where in each event there
is at least one particle with small transverse momentum that does not
pass the final cuts. Since the two samples are disjoint, one can
separately apply \HEJ resummation to each sample and add up the results.

However, implementing the requirement of at least one ``soft'' particle is often
not straightforward with standard fixed-order generators. To
facilitate resummation for the low transverse momentum sample,
\HEJ~2.2 introduces a new option for discarding events in which all
jets are above the analysis threshold. An example is given in section~\ref{sec:low_pt_ex}.

\subsection{Matching to Next-to-Leading Order}
\label{sec:match_NLO}

To improve the accuracy of the obtained total cross section to
next-to-leading order (NLO), one can obviously multiply the \HEJ prediction by a
flat factor of $\sigma_{\text{NLO}} / \sigma_\HEJ$, where
$\sigma_\HEJ$ is the leading-order accurate total cross section
according to \HEJ and $\sigma_{\text{NLO}}$ the total cross section at
NLO. \HEJ 2.2 enables us to achieve NLO accuracy also in
differential distributions. Considering a distribution $d\sigma/d
\mathcal{O}$ in some observable $\mathcal{O}$, we can combine NLO and
\HEJ resummation through the reweighting
\begin{equation}
  \label{eq:match_NLO}
  \left(\frac{d\sigma}{d \mathcal{O}}\right)_{\HEJ + \text{NLO}} =
  \frac{(d\sigma/d\mathcal{O})_{\text{NLO}}}{(d\sigma/d\mathcal{O})_{\HEJ,\text{NLO}}} \left(\frac{d\sigma}{d \mathcal{O}}\right)_{\HEJ}.
\end{equation}
Here, the subscript \HEJ denotes the prediction before reweighting,
\text{NLO} the NLO-accurate prediction, and $\HEJ,$NLO the truncation
of the \HEJ prediction to NLO. In section~\ref{sec:match_NLO_ex}, we
show in an example how to truncate the \HEJ prediction and obtain
NLO-reweighted distributions.

\subsection{Predictions without Fixed-Order Matching}
\label{sec:HEJFOG}

The computational cost for generating the fixed-order input events
rises steeply with the jet multiplicity. For this reason, \HEJ
includes the \HEJFOG, a fast generator based on the leading-order
truncation of the \HEJ matrix element given in
equation~\eqref{eq:ME_fact}. The intended use is that one will generate exact
low-multiplicity input events with a conventional generator and
supplement them with approximate high-multiplicity events using the
\HEJFOG. In \HEJ~2.2, the \HEJFOG includes charged lepton pair
production with jets as a new process. Furthermore, the generation
efficiency for the production of a W boson with jets has been improved
by aligning the rapidity of the W boson with its emitter, reducing the
Monte Carlo uncertainty by a factor of up to 2.


%% file: Application.tex
\section{Example Usage}
\label{sec:usage}

In the following, we show how the new features in \HEJ~2.2 can be used
in practice. For concreteness, we will generate
leading-order events with \textsc{Sherpa}~2.2 and analyse the output with \textsc{Rivet}~3~\cite{Bierlich:2019rhm}. However, we stress that any leading-order
generator that can produce event files in the LHEF
format~\cite{Alwall:2006yp} is supported. In addition to the direct
\textsc{Rivet} interface, \HEJ can write its output to event files in
various formats and allows arbitrary custom analyses via
plugins. Since these options are not new, we refer to the \HEJ
documentation on \url{https://hej.hepforge.org} for details.

\subsection{Same-sign W pair production with jets}
\label{sec:example_WW}

We first consider the process
$pp \to (W^- \to e \bar{\nu}_e) (W^- \to \mu \bar{\nu}_\mu) + \geq 2$
jets with the parameters shown in table~\ref{tab:WW_param}.
$H_T = p_{e\perp} + p_{\bar{\nu}_e\perp} + p_{\mu\perp} +
p_{\bar{\nu}_\mu\perp} + \sum_{i=1}^n p_{i\perp}$ is the sum of the
scalar transverse momenta of all outgoing particles.
\begin{table}[htb]
  \centering
  \renewcommand{\arraystretch}{1.1}
  \begin{tabular}{lp{5mm}l}
    \toprule
    Collider energy && $\sqrt{s} = 13$\,TeV\\[.3em]
    Scales && $\mu_r = \mu_f = \frac{H_T}{2}$\\[.3em]
    PDF set && CT14nlo\\[.3em]
    Electroweak input parameters && $\alpha = 1/132.3572$\\
                    && $m_W = 80.385$\,GeV\\
                    && $\Gamma_W = 2.085$\,GeV\\
                    && $m_Z = 91.1876$\,GeV\\
                    && $\Gamma_Z = 2.4952$\,GeV\\[.3em]
    Jet definition && anti-$k_t$~\cite{Cacciari:2008gp} \\
                    && $R = 0.4$ \\
                    && $p_\perp > 20$\,GeV\\
    \bottomrule
  \end{tabular}
  \caption{Parameters for the production of multiple jets together with a same-sign W boson pair decaying to charged leptons and neutrinos.}
  \label{tab:WW_param}
\end{table}

\subsubsection{Generating leading-order input}
\label{sec:input_WW}

To produce the required leading-order input, we can use \textsc{Sherpa} with the following runcard.
\begin{lstlisting}[language=sherpa,title=\texttt{Run.dat}]
(run){
  EVENTS 10000

  EVENT_OUTPUT LHEF[events_WW2j]

  # theory parameters
  EW_SCHEME 3
  1/ALPHAQED(0) 132.3572
  USE_PDF_ALPHAS 1

  # Z boson mass and width
  MASS[23] 91.1876
  WIDTH[23] 2.4952

  # W boson mass and width
  MASS[24] 80.385
  WIDTH[24] 2.085

  # massless charm quark
  MASSIVE[4] 0
  YUKAWA[4] 0.
  MASS[4] 0.

  # massless bottom quark
  MASSIVE[5] 0
  YUKAWA[5] 0.
  MASS[5] 0.

  # collider beam
  BEAM_ENERGY:=6500.
  BEAM_1 2212 BEAM_ENERGY
  BEAM_2 2212 BEAM_ENERGY

  # PDF
  PDF_LIBRARY LHAPDFSherpa
  PDF_SET CT14nlo

  # Set square of renormalisation and factorisation scale
  SCALES VAR{H_T2/4}

  EVENT_GENERATION_MODE Weighted
  ME_SIGNAL_GENERATOR Comix

  # disable everything beyond fixed order
  FRAGMENTATION Off
  YFS_MODE 0
  MI_HANDLER None
  SHOWER_GENERATOR None
  CSS_MAXEM 0
  BEAM_REMNANTS 0
}(run)

(processes){
  Process 93 93 -> 93 93 11 -12 13 -14
  Order (*,4)
  End process
}(processes)

(selector){
  # require 2 anti-kt jets with pt > 15 GeV and R = 0.4
  FastjetFinder antikt 2 15. 0.0 0.4
}(selector)
\end{lstlisting}
The various settings are explained in more detail in the
\textsc{Sherpa} manual. Since \HEJ treats all quarks as massless, we
have to set the charm and bottom quark masses to zero for
consistency. As explained in section~\ref{sec:low_pt}, leading-order
events containing particles with transverse momenta below the analysis
cuts can still contribute to the resummed prediction. For this reason,
we accept jets with transverse momenta as low as 15\,GeV, despite
having an analysis cut of 20\,GeV. In section~\ref{sec:low_pt_ex}, we
will discuss a computationally more efficient way to incorporate this
contribution from leading-order events that do not pass the analysis
transverse momentum cuts.

We can now generate input events for the case of two jets by running
\begin{lstlisting}[language=sh]
Sherpa
\end{lstlisting}
in the directory containing \texttt{Run.dat}. With the \textsc{Sherpa}
2.2.15 installation inside the \HEJ Docker image this yields a cross
section of $(1.88795 \pm 0.109611)\,$fb. Results may differ between
versions and when re-using integration grids from previous
\textsc{Sherpa} runs. We should then also produce event files for
higher jet multiplicities after adjusting the
\lstinline!EVENT_OUTPUT!, \lstinline!Process!, and
\lstinline!FastjetFinder! entries in \texttt{Run.dat}. Increasing the
number of jets to three leads to a cross section of
$(1.47689 \pm 0.10331)\,$fb.

To avoid the creation of large intermediate event files, we can use
named pipes instead, i.e.~we run \textsc{Sherpa} with
\begin{lstlisting}[language=sh]
mkfifo events_WW2j.lhe
Sherpa &
\end{lstlisting}
Previous \HEJ versions only accepted input from pipes if the total
cross section was equal to the sum of event weights, which is not the
case for \textsc{Sherpa} event files. This restriction is lifted in the new
version $2.2$, which removes a potential bottleneck in time and storage.

\subsubsection{\HEJ resummation}
\label{sec:resum_WW}

In addition to the leading-order event input, \HEJ needs a
configuration file. Adapting the template \texttt{config.yml}
distributed with the \HEJ source code to the parameters listed in
table~\ref{tab:WW_param} we get
\begin{lstlisting}[language=yaml,title=\texttt{config\_WWjets.yml}]
## Number of attempted resummation phase space points for each input event
trials: 10

resummation jets:       # resummation jet properties
  min pt: 20            # minimum jet transverse momentum
  algorithm: antikt     # jet clustering algorithm
  R: 0.4                # jet R parameter

fixed order jets:       # properties of input jets
  min pt: 15
  # by default, algorithm and R are like for resummation jets

## Treatment of the various event classes
## the supported settings are: reweight, keep, discard
## non-resummable events cannot be reweighted
event treatment:
  FKL: reweight # enable LL resummation
  # NLL resummation is not implemented for this process,
  # so we keep all other events as they are
  unordered: keep
  extremal qqbar: keep
  central qqbar: keep
  non-resummable: keep

## Central scale choice or choices
scales: H_T/2

## Analyses
analyses:
  - rivet: [MC_WWINC, MC_WWJETS]  # rivet analysis names
    output: WW2j # name of the yoda files, ".yoda" and scale suffix will be added

## Selection of random number generator and seed
random generator:
  name: mixmax
  seed: 1

## Vacuum expectation value
vev: 246.2196508

## Properties of the weak gauge bosons
particle properties:
  Higgs:
    mass: 125
    width: 0.004165
  W:
    mass: 80.385
    width: 2.085
  Z:
    mass: 91.1876
    width: 2.4952

## Whether or not to include higher order logs
log correction: false
\end{lstlisting}
Here, we choose to pass the resummed events directly to the
\textsc{Rivet} analyses\footnote{The   \texttt{MC\_WWINC} and \texttt{MC\_WWJETS} analyses are written for opposite-sign W boson pair production but can also be used for same-sign W boson pairs.} \texttt{MC\_WWINC} and \texttt{MC\_WWJETS}. Using the Docker
virtualisation software, we can run \HEJ~\cite{docker} with the
following command.
\begin{lstlisting}[language=sh]
docker run -v $PWD:$PWD -w $PWD hejdock/hej HEJ config_WWjets.yml events_WW2j.lhe
\end{lstlisting}
Alternatively, after compiling and installing \HEJ and its dependencies we can use
\begin{lstlisting}[language=sh]
HEJ config_WWjets.yml events_WW2j.lhe
\end{lstlisting}
\HEJ outputs a cross section of $(1.094 \pm 0.06614)\,$fb.
It also produces the \textsc{Rivet} analysis output file
\texttt{WW2j.yoda}. We produce resummed predictions for the
higher-multiplicity event files \texttt{events\_WWnj.lhe} in the same
way, after changing the \lstinline!analyses! entry in
\texttt{config\_WWjets.yml} to
\begin{lstlisting}[language=yaml]
analyses:
  - rivet: [MC_WWINC, MC_WWJETS]
    output: WWnj
\end{lstlisting}
and adjusting the event file name when running \HEJ. To guarantee
statistically independent output, it is also recommended to change the
\lstinline!seed! entry for each run. If we nonetheless use the default
seed with the previously generated event file
\texttt{events\_WW3j.lhe} we obtain a cross section of $(0.905 \pm 0.07534)\,$fb.

\begin{figure}[htb]
  \centering
  \includegraphics[width=0.47\linewidth]{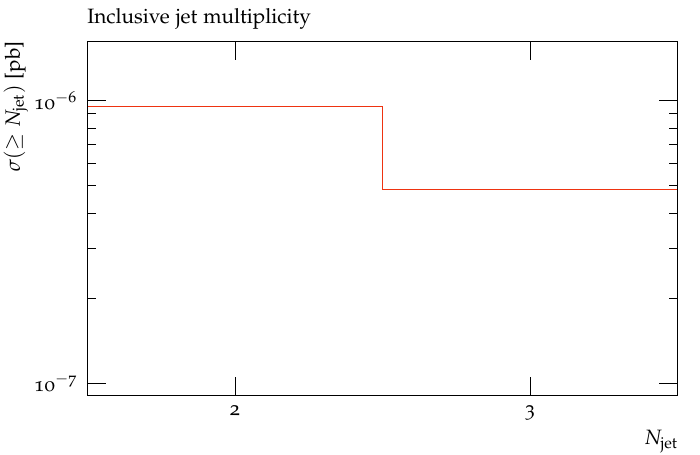}
  \includegraphics[width=0.47\linewidth]{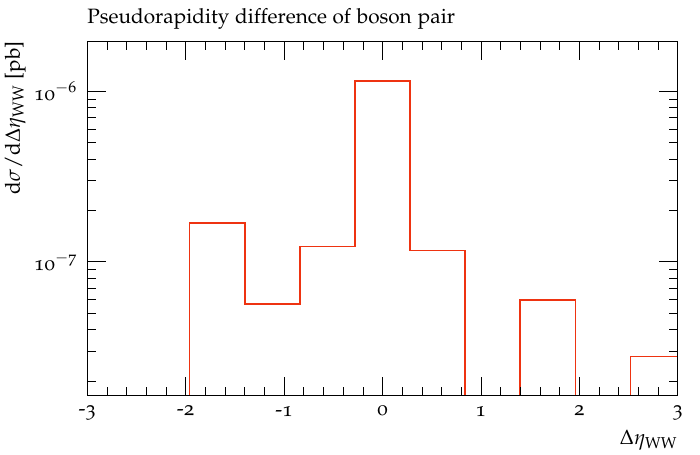}
  \caption{%
    Inclusive $N$-jet cross sections (left) and rapidity difference
    between the two W bosons (right) obtained with \textsc{Sherpa} and
    \HEJ 2.2 for the production of two leptonically decaying W$^-$ bosons
    with at least two jets.
  }
  \label{fig:distr}
\end{figure}

We then combine the results for the different jet multiplicities with
\begin{lstlisting}[language=sh]
yodastack -o WWjets.yoda WW*j.yoda
\end{lstlisting}
and produce plots with
\begin{lstlisting}[language=sh]
rivet-mkhtml WWjets.yoda
\end{lstlisting}
As examples, we show the inclusive jet multiplicities and the
distribution of the rapidity difference between the two W bosons
obtained from resumming fixed-order predictions with two and three
jets in figure~\ref{fig:distr}. In contrast to the recommended usage,
we have not modified the random number generator seeds between runs in
order to facilitate reproducing this figure.\footnote{We have slightly
  changed the plot ranges compared to the default
  \texttt{rivet-mkhtml} settings.} We observe that the bulk of the
events does not pass the analysis cuts.

\subsubsection{Dedicated low transverse momentum runs}
\label{sec:low_pt_ex}

So far, we have generated the leading-order events with significantly
looser transverse momentum cuts than wanted for the final analysis. As
argued in section~\ref{sec:low_pt}, it is more efficient to
split the generation into a high-statistics run with the strict cuts
used in the final analysis and a low-statistics run with loose cuts
where in each leading-order event there is at least one particle that
does not pass the final cuts. For the jet transverse momentum cuts,
this separation is facilitated by a new option in \HEJ 2.2 which
ensures the presence of at least one ``soft'' jet in the input for the
low-statistics run. Note that this option only refers to jets; any
other particles should be generated according to the loose transverse
momentum cuts in both samples.

\begin{figure}[htb]
  \centering
  \includegraphics[width=0.47\linewidth]{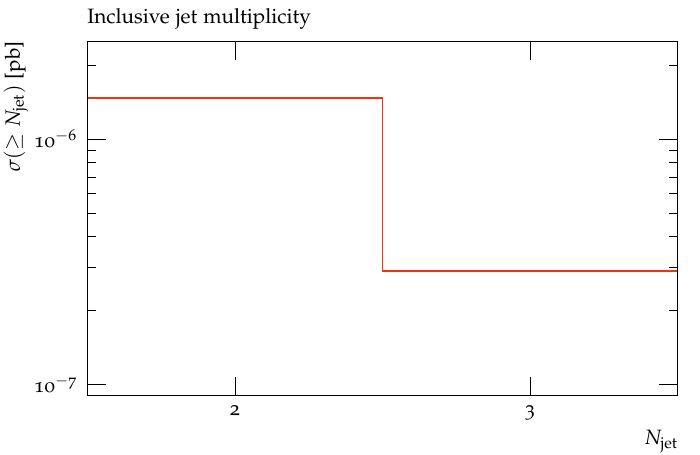}
  \includegraphics[width=0.47\linewidth]{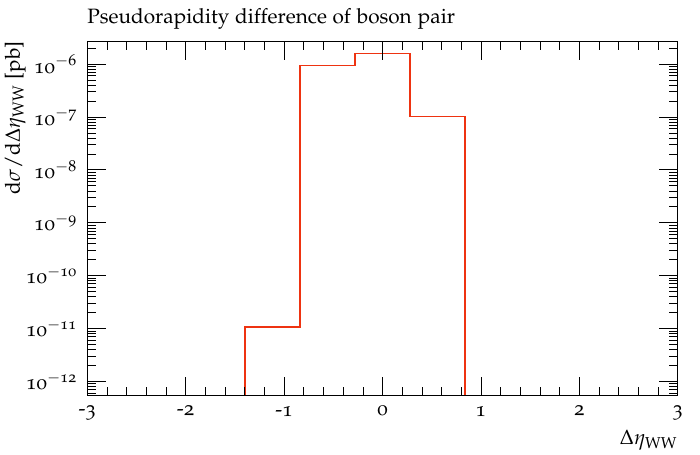}
  \caption{%
    Inclusive $N$-jet cross sections (left) and rapidity difference
    between the two W bosons (right) combining dedicated high and low
    transverse momentum runs.
  }
  \label{fig:distr_lowpt}
\end{figure}

In detail, one should go through the following steps for the present
example of same-sign W pair production with jets:
\begin{enumerate}
\item Generate the high-statistics sample.
\begin{enumerate}
  \item Change the minimum transverse momentum in the \lstinline!FastjetFinder! entry in \texttt{Run.dat} from 15 to 20.
  \item Correspondingly, change
\begin{lstlisting}[language=yaml]
fixed order jets:
  min pt: 15
\end{lstlisting}
    to
\begin{lstlisting}[language=yaml]
fixed order jets:
  min pt: 20
\end{lstlisting}
    in \texttt{config\_WWjets.yml}
  \item Run \textsc{Sherpa} and \HEJ as before.
  \item Revert the changes to both configuration files.
  \end{enumerate}
\item Generate the low-statistics sample.
  \begin{enumerate}
  \item Reduce the number of events generated by Sherpa, for example
by setting the \lstinline!EVENTS! entry in \texttt{Run.dat} to 1000.
  \item Add the entry
\begin{lstlisting}[language=yaml]
require low pt jet: true
\end{lstlisting}
to \texttt{config\_WWjets.yml}.
\item In the \lstinline!event treatment! entry in \texttt{config\_WWjets.yml},
  change all \lstinline!keep! values to \lstinline!discard!. Specifically, change
\begin{lstlisting}[language=yaml]
event treatment:
  FKL: reweight
  unordered: keep
  extremal qqbar: keep
  central qqbar: keep
  non-resummable: keep
\end{lstlisting}
to
\begin{lstlisting}[language=yaml]
event treatment:
  FKL: reweight
  unordered: discard
  extremal qqbar: discard
  central qqbar: discard
  non-resummable: discard
\end{lstlisting}
\item Change the name of the output file, e.g.
\begin{lstlisting}[language=yaml]
analyses:
  - rivet: [MC_WWINC, MC_WWJETS]
    output: WW2j_lowpt
\end{lstlisting}
\item Run \textsc{Sherpa} and \HEJ.
  \end{enumerate}
\end{enumerate}
After repeating these steps for higher jet multiplicities, the
samples can again be combined with
\begin{lstlisting}[language=sh]
yodastack -o WWjets.yoda WW*j.yoda WW*j_lowpt.yoda
\end{lstlisting}
to reproduce the results obtained in section~\ref{sec:resum_WW} with
better statistics. Example plots resulting from
\begin{lstlisting}[language=sh]
rivet-mkhtml WWjets.yoda
\end{lstlisting}
with adjusted axis ranges are shown in figure~\ref{fig:distr_lowpt}.

The cross sections for the various runs are summarised in
table~\ref{tab:xs_ref_lowpt}. Here, ``low $p_\perp$'' refers to the
contribution from the phase-space region where at least one jet is
softer than 20\,GeV. Conversely, ``high $p_\perp$'' implies that all
jets are harder than 20\,GeV. Since we do not impose an upper limit on
the transverse momenta of the jets in the \textsc{Sherpa} runcard, the
low-statistics run covers both of these regions. The ``high
$p_\perp$'' contribution is then removed when running \HEJ with the
\lstinline!require low pt jet! option, which requires that at least
one jet in the fixed-order input is below the analysis cut. While the
contributions from the low transverse momentum run are negligible in
this example, they can become relevant in high-statistics
distributions that probe the phase space near the minimum jet
transverse momentum. What is more, the impact tends to increase with
the jet multiplicity.
\begin{table}[htb]
  \centering
  \begin{tabular}{llcc}
    \toprule
    \multicolumn{2}{l}{Run} &$W^-W^- + 2$ jets&$W^-W^- + 3$ jets\\
    \midrule
    \multirow{2}{*}{\textsc{Sherpa}} & high $p_\perp$ & $1.67517 \pm 0.0702545$ & $1.4744 \pm 0.138443$\\
    & low + high $p_\perp$ & $2.3568 \pm 0.509278$ & $1.70446\pm 0.387036$\\
    \multirow{2}{*}{\HEJ}& high $p_\perp$ & $1.048 \pm 0.05616$ &$1.191 \pm 0.1316$\\
    & low $p_\perp$ $(\times 10^3)$ & $33.07 \pm 42.70$ & $1.018 \pm 1.024$\\
    \bottomrule
  \end{tabular}
  \caption{
    Reference cross sections in femtobarn when using dedicated
    low transverse momentum runs. All runs use the default random
    number generator seeds and the code versions in the \HEJ Docker
    image. \textsc{Sherpa} integration grids (\texttt{Results.db})
    should be deleted between runs to reproduce the exact numbers.
  }
  \label{tab:xs_ref_lowpt}
\end{table}

\subsection{Higgs boson production with one or more jets}
\label{sec:example_H}


We now consider the production of a Higgs boson together with one or
more jets. We use the parameters listed in
table~\ref{tab:H_param}. For the sake of simplicity, we first consider
the limit of an infinitely heavy top quark.

\begin{table}[htb]
  \centering
  \renewcommand{\arraystretch}{1.1}
  \begin{tabular}{lp{5mm}l}
    \toprule
    Collider energy && $\sqrt{s} = 13$\,TeV\\[.3em]
    Scales && $\mu_r = \mu_f = \frac{H_T}{2}$\\[.3em]
    PDF set && CT14nlo\\[.3em]
    Jet definition && anti-$k_t$ \\
                    && $R = 0.4$ \\
                    && $p_\perp > 20$\,GeV\\
    \bottomrule
  \end{tabular}
  \caption{Parameters for the production of a Higgs boson together with at least one jet.}
  \label{tab:H_param}
\end{table}

\begin{figure}[htb]
  \centering
  \includegraphics[width=0.47\linewidth]{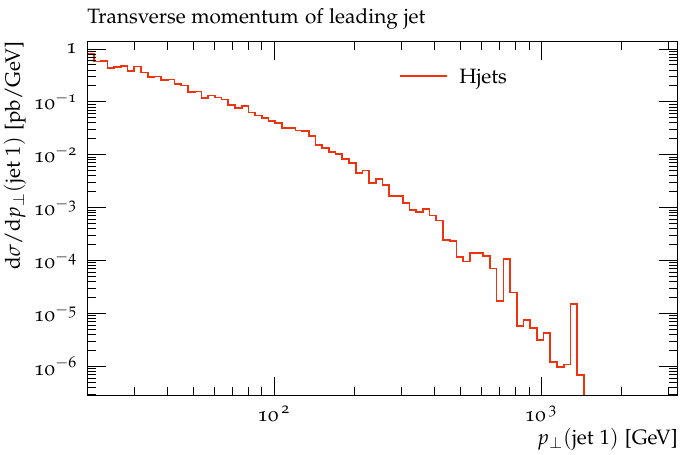}
  \includegraphics[width=0.47\linewidth]{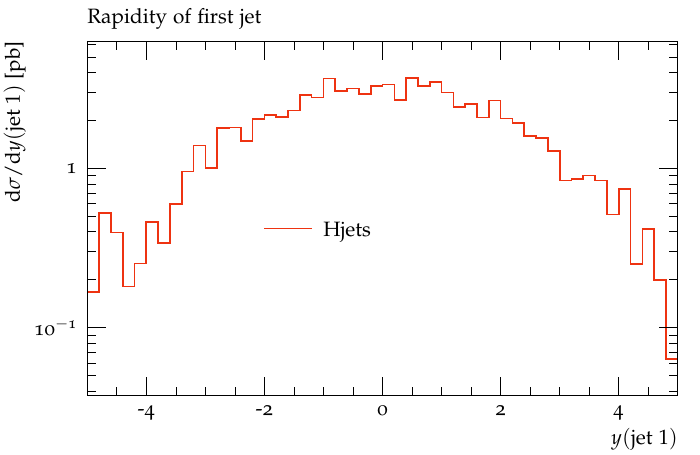}
  \caption{
    Hardest jet transverse momentum and rapidity for the
    production of a Higgs boson together with between one and three jets.
  }
  \label{fig:distr_H}
\end{figure}

In close analogy with section~\ref{sec:example_WW}, we first generate
leading-order input events for Higgs boson production with a single
jet. We use \textsc{Sherpa} with the following run card
\begin{lstlisting}[language=sherpa,title=\texttt{Run.dat}]
(run){
  EVENTS 10000

  EVENT_OUTPUT LHEF[events_H1j]

  # theory parameters
  MODEL HEFT
  USE_PDF_ALPHAS 1
  MASS[25] 125
  WIDTH[25] 0.004165

  # massless charm quark
  MASSIVE[4] 0
  YUKAWA[4] 0.
  MASS[4] 0.

  # massless bottom quark
  MASSIVE[5] 0
  YUKAWA[5] 0.
  MASS[5] 0.

  # collider beam
  BEAM_ENERGY:=6500.
  BEAM_1 2212 BEAM_ENERGY
  BEAM_2 2212 BEAM_ENERGY

  # PDF
  PDF_LIBRARY LHAPDFSherpa
  PDF_SET CT14nlo

  # Set square of renormalisation and factorisation scale
  SCALES VAR{H_T2/4}

  EVENT_GENERATION_MODE Weighted
  ME_SIGNAL_GENERATOR Comix

  # disable everything beyond fixed order
  FRAGMENTATION Off
  YFS_MODE 0
  MI_HANDLER None
  CSS_MAXEM 0
  BEAM_REMNANTS 0
}(run)

(processes){
  Process 93 93 -> 25 93
  Order (*,0,1)
  End process
}(processes)

(selector){
  FastjetFinder antikt 1 15. 0 0.4
}(selector)
\end{lstlisting}
For the resummation, we use a similar \HEJ configuration file as
before. Anticipating further runs with higher multiplicities, we
enable resummation for the supported NLL configurations. In the
present case this concerns configurations involving an unordered
gluon, which first contribute to Higgs boson plus dijet production,
cf.~section~\ref{sec:HEJ_summary}. Since there is no standard
\textsc{Rivet} analysis for stable Higgs boson production, we use the
generic \texttt{MC\_JETS} analysis.
\begin{lstlisting}[language=sherpa,title=\texttt{config\_Hjets.yml}]
trials: 10

resummation jets:
  min pt: 20
  algorithm: antikt
  R: 0.4

fixed order jets:
  min pt: 15

event treatment:
  FKL: reweight
  unordered: reweight
  extremal qqbar: keep
  central qqbar: keep
  non-resummable: keep

scales: H_T/2

analyses:
  - rivet: MC_JETS
    output: H1j

vev: 246.2196508

particle properties:
  Higgs:
    mass: 125
    width: 0.004165
  W:
    mass: 80.385
    width: 2.085
  Z:
    mass: 91.1876
    width: 2.4952

random generator:
  name: mixmax
  seed: 1

log correction: false
\end{lstlisting}
As described in section~\ref{sec:example_WW}, we then add predictions
for higher jet multiplicities. We show reference cross sections in
table~\ref{tab:xs_ref_Hj} and example distributions in
figure~\ref{fig:distr_H}. We stress that, as in most examples shown here,
the \textsc{Sherpa} and \HEJ cross sections are not directly comparable.
This is because the \textsc{Sherpa} runcards use a lower transverse momentum
cut than \HEJ, for the reason explained in section~\ref{sec:low_pt}.
Therefore a significant fraction of the input events will not pass the
cuts after resummation (see section~\ref{sec:low_pt_ex} for an example
on how to deal with this in a more efficient way).

\begin{table}[htb]
  \centering
  \begin{tabular}{lccc}
    \toprule
    Run & $H+1$ jet& $H+2$ jets& $H+3$ jets\\
    \midrule
    \textsc{Sherpa}&$24.25 \pm 0.494414$ & $14.0615 \pm 0.469809$ & $6.38016 \pm 0.43697$\\
    \HEJ&$11.20 \pm 0.3732$ & $4.420 \pm 0.1953$ & $1.657\pm 0.1558$\\
    \bottomrule
  \end{tabular}
  \caption{Reference cross sections in picobarn for Higgs boson plus jet production.}
  \label{tab:xs_ref_Hj}
\end{table}

\subsubsection{Quark mass corrections}
\label{sec:H_mt}

For accurate predictions in the high-energy region, we have to take
into account the finite top quark mass. In the following, we assume a
mass of 174\,GeV. On the \textsc{Sherpa} side, we can add \textsc{AMEGIC}~\cite{Krauss:2001iv}
 and \textsc{OpenLoops}~\cite{Cascioli:2011va,Buccioni:2019sur} to
\lstinline!ME_SIGNAL_GENERATOR! and insert the following lines into the
\lstinline[language=sherpa]!(run)! block:
\begin{lstlisting}[language=sherpa]
  # finite top mass effects
  KFACTOR GGH
  OL_IGNORE_MODEL 1
  OL_PARAMETERS preset 2 allowed_libs pph2,pphj2,pphjj2 psp_tolerance 1.0e-7
\end{lstlisting}
\HEJ needs to be compiled with support for
\textsc{QCDLoop}~\cite{Carrazza:2016gav} to incorporate quark mass
corrections in Higgs boson production. We can include them by adding
\begin{lstlisting}[language=yaml]
Higgs coupling:
   use impact factors: false
   mt: 174
\end{lstlisting}
to the configuration file.

For higher jet multiplicities, we face the problem that it is no
longer feasible to compute the leading-order input with exact
dependence on the top quark mass $m_t$. However, we can still retain this
dependence and also include the dependence on the bottom-quark mass $m_b$ in
the high-energy resummation.

As in equation~(\ref{eq:weight}), the weight $w$ of a leading-order matched
resummation event has the following dependence on the leading-order
matrix element $\mathcal{M}_{\text{LO}}$ and the all-order \HEJ matrix
element $\mathcal{M}_{\HEJ}(m_b, m_t)$:
\begin{equation}
  \label{eq:ME_wt}
  w \propto  \frac{\lvert\mathcal{M}_{\text{LO}}(m_b, m_t)\rvert^2 \lvert\mathcal{M}_{\HEJ}(m_b, m_t)\rvert^2}{\lvert\mathcal{M}_{\HEJ, \text{LO}}(m_b, m_t)\rvert^2}.
\end{equation}
For consistency, the values for the quark
masses have to match those used in
$\mathcal{M}_{\text{LO}}$. Therefore, if the leading-order input is
only known for $m_b \to 0, m_t \to \infty$ the correct reweighting
factor is
\begin{equation}
  \label{eq:ME_wt_mtinf}
  w \propto  \frac{\lvert\mathcal{M}_{\text{LO}}(0, \infty)\rvert^2 \lvert\mathcal{M}_{\HEJ}(m_b, m_t)\rvert^2}{\lvert\mathcal{M}_{\HEJ, \text{LO}}(0, \infty)\rvert^2} =  \frac{\lvert\mathcal{M}_{\text{LO}}(0, \infty)\rvert^2 \lvert\mathcal{M}_{\HEJ}(m_b, m_t)\rvert^2}{\lvert\mathcal{M}_{\HEJ, \text{LO}}(m_b, m_t)\rvert^2} \times \frac{\lvert\mathcal{M}_{\HEJ, \text{LO}}(m_b, m_t)\rvert^2}{\lvert\mathcal{M}_{\HEJ, \text{LO}}(0, \infty)\rvert^2}.
\end{equation}
Currently, there is no built-in \HEJ option for choosing different
quark mass values in $\mathcal{M}_{\HEJ}$ and $\mathcal{M}_{\HEJ,
\text{LO}}$. However, \HEJ supports flexible custom analyses, which
allow us to manually reweight by the correction factor
$\lvert\mathcal{M}_{\HEJ, \text{LO}}(m_b, m_t)\rvert^2 / \lvert\mathcal{M}_{\HEJ, \text{LO}}(0,
\infty)\rvert^2$ in equation~\eqref{eq:ME_wt_mtinf}. We can also use this
opportunity to include bottom quark mass corrections in the case where
only the exact leading-order dependence on the top quark mass is available.

\begin{figure}[htb]
  \centering
  \includegraphics[width=0.47\linewidth]{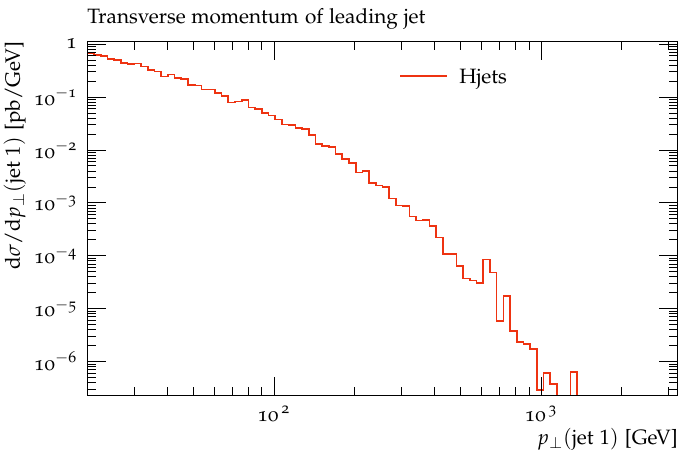}
  \includegraphics[width=0.47\linewidth]{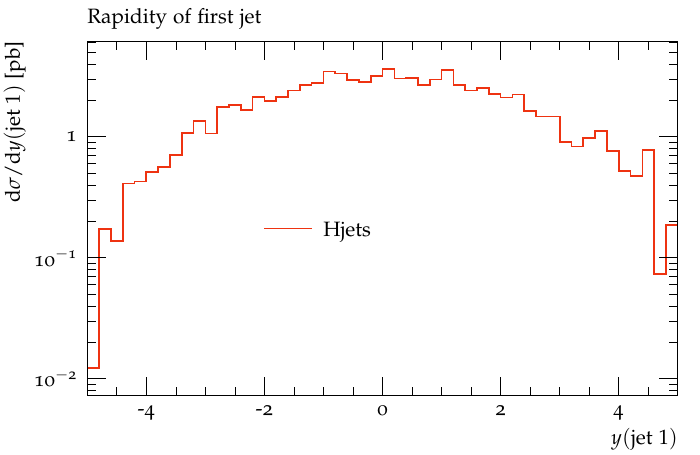}
  \caption{
    Hardest jet transverse momentum and rapidity for the
    production of a Higgs boson together with between one and three
    jets including finite quark mass effects.
  }
  \label{fig:distr_H_mtmb}
\end{figure}

Custom analyses are described in the \HEJ user documentation on
\url{https://hej.hepforge.org}, where also a template is provided. The
reweighting can be implemented as shown here:
\begin{lstlisting}[language=C++,title=\texttt{higgs\_matching\_analysis.cc}]
#include <string>
#include <memory>

#include "HEJ/Analysis.hh"
#include "HEJ/Config.hh"
#include "HEJ/event_types.hh"
#include "HEJ/Event.hh"
#include "HEJ/HiggsCouplingSettings.hh"
#include "HEJ/MatrixElement.hh"
#include "HEJ/RivetAnalysis.hh"
#include "HEJ/YAMLreader.hh"

#include "yaml-cpp/yaml.h"

namespace LHEF {
  class HEPRUP;
}

namespace {
  // Dummy function for alpha_s(mu)
  // Since the dependence on alpha_s cancels out in the matrix element ratio
  // we can return any (finite, non-zero) value we like
  double alpha_s_dummy(double /* mu */) {
    return 1.;
  }

  // Helper function to initialise modulus square of HEJ matrix element
  HEJ::MatrixElement create_HEJ_ME(
    YAML::Node const & yaml_config,
    std::string const &  Higgs_coupling_name
  ) {
      HEJ::MatrixElementConfig config;
      config.log_correction = false;
      config.ew_parameters = HEJ::get_ew_parameters(yaml_config);
      config.Higgs_coupling = HEJ::get_Higgs_coupling(
        yaml_config,
        Higgs_coupling_name
      );
      return HEJ::MatrixElement{alpha_s_dummy, config};
  }

  // Check if we perform resummation for an event
  bool is_resummable(HEJ::Event const & event) {
    switch(event.type()) {
    case HEJ::event_type::FKL:
    case HEJ::event_type::unordered_backward:
    case HEJ::event_type::unordered_forward:
      return true;
    default:
      return false;
    }
  }

  // Our custom analysis
  class HiggsMatchingAnalysis: public HEJ::Analysis {
    public:
      HiggsMatchingAnalysis(
        YAML::Node const & config,
        LHEF::HEPRUP const & heprup
      ):
        me_exact_{create_HEJ_ME(config, "Exact Higgs coupling")},
        me_approx_{create_HEJ_ME(config, "LO Higgs coupling")},
        rivet_analysis_{config["Rivet analysis"], heprup}
    {}

      void fill(
        HEJ::Event const & ev,
        HEJ::Event const & LO_event
      ) override {
        auto event = ev;
        if(is_resummable(event)) {
          // Compute ratio of exact and approximate matrix element squares
          // truncated to leading order.
          // In HEJ, we split the squares of tree-level matrix elements into a
          // scale-dependent "parametric" factor containing the couplings and a
          // scale-independent "kinetic" factor. Since the dependence on the
          // couplings cancels exactly in the ratio, we only need the latter
          // part `tree_kin` here.
          // To account for the possibility of interference, `tree_kin` returns
          // a `std::vector` instead of a single value. Here, the `std::vector`
          // has only a single element.
          const double me_exact = me_exact_.tree_kin(LO_event).front();
          const double me_approx = me_approx_.tree_kin(LO_event).front();
          const double reweight = me_exact / me_approx;
          event.central().weight *=  reweight;
          // If we perform scale variation we also have to rescale all other
          // weights
          for(auto & var: event.variations()) {
            var.weight *= reweight;
          }
        }
        // Pass the potentially reweighted event to Rivet
        rivet_analysis_.fill(event, LO_event);
      }

      bool pass_cuts(
        HEJ::Event const & /* event */,
        HEJ::Event const & /* LO_event */
      ) override {
        return true;
      }

      void set_xs_scale(double scale) override {
        rivet_analysis_.set_xs_scale(scale);
      }

      void finalise() override {
        rivet_analysis_.finalise();
      }

  private:
    HEJ::MatrixElement me_exact_;
    HEJ::MatrixElement me_approx_;
    HEJ::RivetAnalysis rivet_analysis_;
  };
}

extern "C"
__attribute__((visibility("default")))
std::unique_ptr<HEJ::Analysis> make_analysis(
    YAML::Node const & config, LHEF::HEPRUP const & heprup
){
  return std::make_unique<HiggsMatchingAnalysis>(config, heprup);
}
\end{lstlisting}
We can then compile the analysis into a shared object library with a
compiler supporting C++17, for instance a recent version of
\texttt{g++}:
\begin{lstlisting}[language=sh]
g++ -Wall -Wextra $(HEJ-config --cxxflags) -fPIC -shared -O2 \
  -fvisibility=hidden \
  -Wl,-soname,libhiggs_matching_analysis.so \
  -o libhiggs_matching_analysis.so higgs_matching_analysis.cc
\end{lstlisting}
To use our custom analysis, we adjust the \HEJ configuration file. We
use YAML anchors (starting with \lstinline!&!) and references
(starting with \lstinline!*!) to ensure that the settings passed to
the analysis are consistent. The following code listing is for the
case of a leading-order prediction in the infinite top-quark mass
limit.
\begin{lstlisting}[language=yaml,title=\texttt{config\_Hjets\_mbmt.yml}]
trials: 10

resummation jets:
  min pt: 20
  algorithm: antikt
  R: 0.4

fixed order jets:
  min pt: 15

event treatment:
  FKL: reweight
  unordered: reweight
  extremal qqbar: keep
  central qqbar: keep
  non-resummable: keep

scales: H_T/2

vev: 246.2196508

particle properties: &particle_properties
  Higgs:
    mass: 125
    width: 0.004165
  W:
    mass: 80.385
    width: 2.085
  Z:
    mass: 91.1876
    width: 2.4952

random generator:
  name: mixmax
  seed: 1

Higgs coupling: &exact_h_coupling
   use impact factors: false
   mt: 174
   include bottom: true
   mb: 2.8

analyses:
  - plugin: ./libhiggs_matching_analysis.so
    Exact Higgs coupling: *exact_h_coupling
    LO Higgs coupling:
      use impact factors: true
    Rivet analysis:
      rivet: MC_JETS
      output: H1j
    vev: 246.2196508
    particle properties: *particle_properties

log correction: false
\end{lstlisting}
If the leading order prediction includes the exact dependence on the
top-quark mass, but not the dependence on the bottom-quark mass, one
should replace the \lstinline!LO Higgs coupling! entry by
\begin{lstlisting}[language=yaml]
    LO Higgs coupling:
      use impact factors: false
      mt: 174
\end{lstlisting}
Table \ref{tab:xs_ref_Hj_mtmb} lists reference cross sections and
figure~\ref{fig:distr_H_mtmb} shows corresponding distributions
combining samples with up to three jets.

\begin{table}[htb]
  \centering
  \begin{tabular}{lccc}
    \toprule
    Run & $H+1$ jet&  $H+2$ jets & $H+3$ jets\\
    \midrule
    \textsc{Sherpa} & $25.911 \pm 0.442621$ &see tab.~\ref{tab:xs_ref_Hj} & see tab.~\ref{tab:xs_ref_Hj}\\
\HEJ &$11.50 \pm 0.2616$&$4.409 \pm 0.1947$&$1.648 \pm 0.1544$\\
    \bottomrule
  \end{tabular}
  \caption{
    Reference cross sections in picobarn for Higgs boson plus
    jets production with finite quark mass effects. All \HEJ resummation
    results account for massive bottom and top quarks. The
    \textsc{Sherpa} prediction for $H+1$ jet uses the physical
    top-quark mass, but assumes a massless bottom quark. The
    fixed-order predictions for higher multiplicities do not include
    finite quark mass effects.
  }
  \label{tab:xs_ref_Hj_mtmb}
\end{table}

\subsubsection{Matching to next-to-leading order}
\label{sec:match_NLO_ex}

Using \HEJ 2.2, we can extend the fixed-order accuracy of
distributions from LO to NLO, cf.~section~\ref{sec:match_NLO}. In the
following, we consider NLO matching for the production of a Higgs boson
together with at least one jet. While we derive the matching in the
limit of an infinitely heavy top quark, the resulting rescaling factors
can also be used to improve HEJ predictions incorporating quark-mass
corrections. We apply the matching bin-by-bin in the resulting
histograms. We can generate an NLO prediction using \textsc{Sherpa} and
\textsc{OpenLoops} by adjusting the run card as follows.
\begin{lstlisting}[language=sherpa,title=\texttt{Run.dat}]
(run){
  EVENTS 10000

  MODEL HEFT
  USE_PDF_ALPHAS 1
  MASS[25] 125
  WIDTH[25] 0.004165

  MASSIVE[4] 0
  YUKAWA[4] 0.
  MASS[4] 0.

  MASSIVE[5] 0
  YUKAWA[5] 0.
  MASS[5] 0.

  BEAM_ENERGY:=6500.
  BEAM_1 2212 BEAM_ENERGY
  BEAM_2 2212 BEAM_ENERGY

  PDF_LIBRARY LHAPDFSherpa
  PDF_SET CT14nlo

  SCALES VAR{H_T2/4}

  EVENT_GENERATION_MODE Weighted
  ME_SIGNAL_GENERATOR Comix OpenLoops

  FRAGMENTATION Off
  YFS_MODE 0
  MI_HANDLER None
  CSS_MAXEM 0
  BEAM_REMNANTS 0

  ANALYSIS Rivet
  ANALYSIS_OUTPUT Hj_NLO
}(run)

(processes){
  Process 93 93 -> 25 93
  Order (*,0,1)
  NLO_QCD_Mode Fixed_Order
  NLO_QCD_Part BVIRS
  Integration_Error 0.02
  Loop_Generator OpenLoops
  End process
}(processes)

(selector){
  FastjetFinder antikt 1 20. 0 0.4
}(selector)

(analysis){
  BEGIN_RIVET {
  USE_HEPMC_SHORT 1
  IGNOREBEAMS 1
  -a MC_JETS
 } END_RIVET
}(analysis)
\end{lstlisting}
The resulting cross section is $(18.4425 \pm 3.83924)\,$pb.

\begin{figure}[htb]
  \centering
  \includegraphics[width=0.47\linewidth]{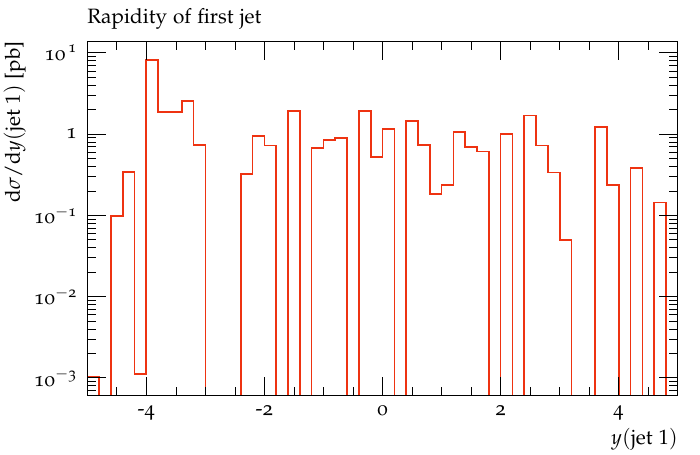}
  \caption{%
    Differential NLO matching for the rapidity distribution of the
    hardest jet produced together with a Higgs boson.
  }
  \label{fig:distr_H_NLO}
\end{figure}

For the \HEJ prediction, we add the option
\begin{lstlisting}[language=yaml]
NLO truncation:
  enabled: true
  nlo order: 1    # number of jets
\end{lstlisting}
to the configuration file \texttt{config\_Hjets.yml} from
section~\ref{sec:example_H}, change the name of the \textsc{Rivet}
output file to \texttt{Hj\_HEJ\_NLO\_1j.yoda}, and run \HEJ on the
previously generated leading-order input file assuming an infinitely
heavy top quark. We obtain a cross section of $(7.846 \pm
1.091)$\,pb. The result is \emph{exclusive} in the number of jets; in
contrast to the \textsc{Sherpa} prediction the contribution from
events with two jets is not included, yet. To generate the missing
piece, we remove the NLO process configuration from the above
\textsc{Sherpa} run card, i.e.\ we change the
\lstinline!ANALYSIS_OUTPUT! to \texttt{Hjj\_LO}, remove
\textsc{OpenLoops} from \texttt{ME\_SIGNAL\_GENERATOR} and delete the
following lines.  {\color{gray}
\begin{lstlisting}
  NLO_QCD_Mode Fixed_Order
  NLO_QCD_Part BVIRS
  Integration_Error 0.02
  Loop_Generator OpenLoops
\end{lstlisting}
}
We then increase the number of jets to two and run \textsc{Sherpa}
to generate a file \texttt{Hjj\_LO.yoda}, with a total cross section
of $(8.80745 \pm 0.321808)$\,pb. We add this contribution to the
truncated \HEJ result:
\begin{lstlisting}[language=sh]
  yodastack -o Hj_HEJ_NLO.yoda Hj_HEJ_NLO_1j.yoda Hjj_LO.yoda
\end{lstlisting}
To obtain the final NLO matched prediction, we should multiply each
histogram in the original \HEJ output by the ratio of the
corresponding histograms in \texttt{Hj\_NLO.yoda} and
\texttt{Hj\_HEJ\_NLO.yoda}. The following script gives an example of
how this reweighting can be implemented. For the sake of brevity we
have omitted the error handling code, which is of course essential in
actual applications.
\begin{lstlisting}[language=python,title=\texttt{reweight\_NLO.py}]
#!/usr/bin/env python

import yoda

nlo = yoda.read("Hj_NLO.yoda")
hej_nlo = yoda.read("Hj_HEJ_NLO.yoda")
full = yoda.read("Hjets.yoda")
full_rescaled = {}

for path, full in full.items():
    try:
        reweight_fact_scatter = hej_nlo[path] / nlo[path]
        reweight_fact = yoda.core.Histo1D(path, full.title())
        reweight_fact.addBins(full.xEdges())
        for p in reweight_fact_scatter.points():
            reweight_fact.fill(p.x(), p.y())
        full_rescaled[path] = full / reweight_fact
        full_rescaled[path].setPath(path)
    except:
        pass

yoda.write(full_rescaled, "Hjets_rescaled.yoda")
\end{lstlisting}
Generating \texttt{Hjets.yoda} as described in section~\ref{sec:H_mt},
the resulting NLO-matched rapidity distribution of the hardest jet is
shown in figure~\ref{fig:distr_H_NLO}.  Note that the
\texttt{MC\_JETS} \textsc{Rivet} analysis employs adaptive binning
for a number of distributions, causing the division of the respective
histograms to fail. This problem can of course be circumvented by
using a custom analysis.

\subsection{Charged lepton pair production with many jets}
\label{sec:HEJFOG_ex}

The production of two charged leptons with jets was first implemented
in \HEJ~2.1~\cite{Andersen:2021qma} for at most moderate jet
multiplicities, where exact fixed-order matching is feasible. To
overcome this difficultly, in \HEJ~2.2 approximate high-multiplicity
events can be generated with the \HEJ Fixed Order Generator.

In the following example, we consider the process $pp \to \mu^+ \mu^-
+ \geq$ 2 jets with the same parameters as in previous examples, see
table~\ref{tab:WW_param}. We assume that predictions including up to 4
jets have already been produced as described in
sections~\ref{sec:example_WW} and~\ref{sec:example_H}. To generate
input events with 5 jets, we adapt the configuration file
\texttt{configFO.yml} distributed together with \HEJ:
\begin{lstlisting}[language=yaml,title=\texttt{configFO\_Zjets.yml}]
## Number of generated events
events: 10000

jets:
  min pt: 15        # minimal jet pt, should be slightly below analysis cut
  peak pt: 20       # peak pt of jets, should be at analysis cut
  algorithm: antikt # jet clustering algorithm
  R: 0.4            # jet R parameter
  max rapidity: 5   # maximum jet rapidity

## Particle beam
beam:
  energy: 6500
  particles: [p, p]

## PDF ID according to LHAPDF
pdf: 13100

## Scattering process
process: p p => mu mu_bar 5j

## Fraction of events with two extremal emissions in one direction
## that contain an subleading emission e.g. unordered emission
subleading fraction: 0.05

## Allow different subleading configurations.
## By default all subleading channels are disabled.
subleading channels:
  - unordered

## Unweighting parameters
## remove to obtain weighted events
unweight:
  sample size: 10000  # should be similar to "events:", but not more than ~10000
  max deviation: 0

## Central scale choice
scales: H_T / 2

## Event output files
event output:
  - events_Z5j.lhe

## Selection of random number generator and seed
random generator:
  name: mixmax
  seed: 1

## Vacuum expectation value
vev: 246.2196508

## Properties of the weak gauge bosons
particle properties:
  Higgs:
    mass: 125
    width: 0.004165
  W:
    mass: 80.385
    width: 2.085
  Z:
    mass: 91.1876
    width: 2.4952
\end{lstlisting}
By setting the \lstinline!peak pt! option we ensure that most events
are generated above the analysis cut of 20\,GeV. This means that there
is no need for two separate runs with different transverse momentum
cuts as described in section~\ref{sec:low_pt_ex} for conventional
fixed-order generators. We can now generate events with
\begin{lstlisting}
docker run -v $PWD:$PWD -w $PWD hejdock/hej HEJFOG configFO_Zjets.yml
\end{lstlisting}
when using the \HEJ Docker container or
\begin{lstlisting}
HEJFOG configFO_Zjets.yml
\end{lstlisting}
when using a local \HEJ installation. We obtain a cross section of
$(11.53 \pm 1.607)\,$pb. Afterwards, we can run \HEJ on the resulting
\texttt{events\_Z5j.lhe} event file to obtain resummed predictions
for $pp \to \mu^+ \mu^- +$ 5 jets. These could then be combined
with the results for lower multiplicities as described in
section~\ref{sec:resum_WW}.


%% file: Conclusions.tex
\section{Conclusions}
\label{sec:conclusions}

The \HEJ Monte Carlo event generator provides accurate high-energy
descriptions for a steadily growing range of scattering
processes. The new version 2.2 adds predictions for the QCD corrections to the
production of two W bosons with the same charge together with two or
more jets, which are pivotal for experimental measurements of weak
boson fusion. A further major improvement concerns the gluon-fusion
production of a Higgs boson with jets, where for the first time final
states with only a single jet are included in the description.

While LO-matched predictions for charged lepton pair production with
jets were already available in \HEJ 2.1, the new release allows to
supplement these with LL-accurate high-energy corrections for high
multiplicities where exact LO generation may no longer be
feasible. Furthermore, \HEJ 2.2 allows NLO matching at the level
of differential distributions and facilitates a computationally more efficient
event input generation by separating off low-$p_\perp$ events with a
small contribution to the final predictions.

We have given detailed examples for the new features. The program code as
well as comprehensive documentation including options added in
previous versions are available on \url{https://hej.hepforge.org}.

\section*{Acknowledgements}
\label{sec:acknowledgements}

\HEJ uses \texttt{FastJet}~\cite{Cacciari:2011ma} and \texttt{LHAPDF}~\cite{Buckley:2014ana}.

We are grateful to the other members of the \HEJ collaboration for helpful
comments and discussions.   We are pleased to acknowledge funding from the UK Science and
Technology Facilities Council, the Royal Society, and the ERC Starting
Grant 715049 ``QCDforfuture''.  For the purpose of open access, the authors have
applied a Creative Commons Attribution (CC BY) licence to any Author Accepted
Manuscript version arising from this submission.


%% file: Installation.tex
\newcommand{\fakehref}[2]{#2} 

\section{Download and installation}
\label{sec:installation}

Detailed documentation for \HEJ 2.2 can be found on
\url{https://hej.hepforge.org/}. In the following, we briefly explain
how to download and install the program.

\hypertarget{download}{%
\subsection{Download}\label{download}}

A tar archive of the \HEJ 2 source code can be downloaded and
decompressed with the command:
\begin{lstlisting}
curl https://hej.hepforge.org/downloads?f=HEJ_2.2.tar.gz | tar -xz
\end{lstlisting}
To obtain the latest stable \HEJ version, \lstinline!HEJ_2.2.tar.gz! should be
replaced by \lstinline!HEJ.tar.gz!.

Alternatively, the \HEJ source code can be obtained by installing the
\fakehref{https://git-scm.com/}{git version control system}~\cite{git} and running:
\begin{lstlisting}
git clone https://phab.hepforge.org/source/hej.git
\end{lstlisting}

We also provide a \fakehref{https://hub.docker.com/r/hejdock/hej}{Docker
image} containing a \HEJ 2 installation (including the HEJ Fixed Order
Generator). This image can be pulled with:
\begin{lstlisting}
docker pull hejdock/hej
\end{lstlisting}
When using the Docker image the remaining installation steps can be
skipped.

In addition to working with Docker, these images will also work with
\fakehref{https://apptainer.org/}{Apptainer}/\fakehref{https://docs.sylabs.io/guides/latest/user-guide/}{SingularityCE}~\cite{apptainer,singularity}.
This comes with a caveat that you may need to source the software inside
the image before running \HEJ:
\begin{lstlisting}
source /cvmfs/pheno.egi.eu/HEJV2/HEJ_env.sh
\end{lstlisting}

Users can also make use of the MPI libraries included in the images for
efficient fixed order event generation for \HEJ input using Sherpa which
is configured with MPI support. Note that \HEJ itself does not provide
support for MPI.

\hypertarget{prerequisites}{%
\subsection{Prerequisites}\label{prerequisites}}

Before installing \HEJ 2, you need the following programs and libraries:
\begin{itemize}
\item
  \fakehref{https://cmake.org/}{CMake}~\cite{cmake} version 3.1
\item
  A compiler supporting the C++17 standard, for example
  \fakehref{https://gcc.gnu.org/}{gcc} 7 or later~\cite{gcc}
\item
  \fakehref{http://fastjet.fr/}{FastJet}~\cite{Cacciari:2011ma}
\item
  \fakehref{https://gitlab.cern.ch/CLHEP/CLHEP}{CLHEP}~\cite{CLHEP} version 2.3
\item
  \fakehref{https://lhapdf.hepforge.org/}{LHAPDF}~\cite{Buckley:2014ana} version 6
\item
  The {IOStreams} and {uBLAS} \fakehref{https://www.boost.org}{boost}~\cite{boost}
  libraries
\item
  \fakehref{https://github.com/jbeder/yaml-cpp}{yaml-cpp}~\cite{yaml-cpp}
\item
  \texttt{autoconf} and \texttt{automake} for
  \fakehref{https://github.com/vermaseren/form}{FORM}~\cite{Vermaseren:2000nd}
\end{itemize}

In addition, some optional features have additional dependencies:
\begin{itemize}
\item \fakehref{https://github.com/scarrazza/qcdloop}{Version 2 of
    QCDLoop}~\cite{Carrazza:2016gav} is required to include finite top
  mass corrections in Higgs boson + jets production.
\item
  \fakehref{https://hepmc.web.cern.ch/hepmc/}{HepMC~\cite{Dobbs:2001ck} versions 2 and 3} enable
  event output in the respective format.
\item
  \fakehref{https://rivet.hepforge.org/}{Rivet}~\cite{Bierlich:2019rhm} version 3.1.4 or later
  together with HepMC 2 or 3 allow using Rivet analyses.
\item
  \fakehref{https://github.com/BlueBrain/HighFive}{HighFive}\cite{highfive}
  has to be installed in order to read and write event files in the
  \fakehref{https://www.hdfgroup.org/}{HDF5}~\cite{hdf5}-based format
  suggested in \cite{Hoche:2019flt}.
\end{itemize}

We strongly recommend to install these programs and libraries to
standard locations:
\begin{itemize}
\item
  The executable files should be inside one of the directories listed in
  the \lstinline!PATH! environment variable. This concerns \texttt{cmake}, the C++
  compiler, and the executables contained in \texttt{autoconf} and \texttt{automake}.
\item
  The library header files ending with \texttt{.h}, \texttt{.hh}, or \texttt{.hpp} should be
  in a directory where they are found by the C++ compiler. For \texttt{gcc} or
  \texttt{clang}, custom locations can be specified using the
  \lstinline!CPLUS_INCLUDE_PATH! environment variable.
\item The compiled library files ending with \texttt{.a},
  \texttt{.so}, or \texttt{.dylib} should be in a directory where they
  are found by the linker. Custom locations can be set via the
  \lstinline!LIBRARY_PATH! environment variable. For shared object
  libraries (\texttt{.so} or \texttt{.dylib}) custom locations should
  also be part of \lstinline!LD_LIBRARY_PATH! on linux and
  \lstinline!DYLD_FALLBACK_LIBRARY_PATH! or
  \lstinline!DYLD_LIBRARY_PATH! on macOS.
\end{itemize}

\hypertarget{compilation}{%
\subsection{Compilation}\label{compilation}}

To compile and install \HEJ 2 run:
\begin{lstlisting}
cmake source/directory -DCMAKE_INSTALL_PREFIX=target/directory
make install
\end{lstlisting}
\texttt{source/directory} is the directory containing the file
\texttt{CMakeLists.txt}. If you omit
\lstinline!-DCMAKE_INSTALL_PREFIX=target/directory!, \HEJ 2 will be
installed to some default location.

In case some of the aforementioned prerequisites are not found by
\texttt{cmake} you can give a hint by adding an additional argument
\texttt{-Dlibname\_ROOT\_DIR=/directory/with/library}, where
\texttt{libname} should be replaced by the name of the library in
question. For example, if {FastJet} is installed in the subdirectory
\texttt{.local} of your home directory with the \texttt{libfastjet.*} library files in
\texttt{.local/lib} and the header files ending with \texttt{.hh} in
\texttt{.local/include/fastjet} you can pass
\lstinline!-Dfastjet_ROOT_DIR=$HOME/.local! to \texttt{cmake}.

If \texttt{cmake} fails to find (the correct) boost path, try setting
\lstinline!-DBOOST_ROOT=/path/to/boost!, this will force \texttt{cmake} to
search for boost only in \texttt{/path/to/boost}.

To not include specific packages one can add
\lstinline!-DEXCLUDE_packagename=TRUE! to \texttt{cmake}, e.g. by setting
\lstinline!-DEXCLUDE_rivet=TRUE! \HEJ 2 will not be interfaced to Rivet
even if it is available on the system.

\hypertarget{testing}{%
\subsection{Testing}\label{testing}}

To test your installation, download the NNPDF 2.3 PDF set with:
\begin{lstlisting}
lhapdf install NNPDF23_nlo_as_0119
\end{lstlisting}
and run:
\begin{lstlisting}
make test
\end{lstlisting}
The test data of \HEJ are stored in a
\fakehref{https://git-lfs.github.com/}{Git Large File Storage}~\cite{git-lfs} format.
\lstinline!git clone! therefore requires \texttt{git-lfs} to download the
data correctly.

\subsection{Running}

To run \HEJ, use the command
\begin{lstlisting}
HEJ config.yml events.lhe
\end{lstlisting}
where \texttt{config.yml} is a configuration file and
\texttt{events.lhe} a Les Houches event
file~\cite{Alwall:2006yp}. Explicit examples are given in
section~\ref{sec:usage}.
